\documentclass[aps,prd,reprint,showpacs,article,floatfix,nofootinbib]{revtex4-1}
\pdfoutput=1

\usepackage{fullpage}
\usepackage{amsmath}
\usepackage{amssymb}
\usepackage{graphicx}
\usepackage{color,soul}
\usepackage{xcolor}
\usepackage{subcaption}
\usepackage{hyperref}

\begin{document}

\title{Specific Heats for  
Quantum BTZ Black Holes in Extended Thermodynamics}

\author{Clifford V. Johnson$^\dagger$ and Roberto Nazario$^*$\\}
\affiliation{$^\dagger$Department of Physics, University of
 California, 
 Santa Barbara, CA 93106-9530, U.S.A.}
\affiliation{
$^*$Department of Physics and Astronomy, University of
Southern California,
 Los Angeles, CA 90089-0484, U.S.A.}


\begin{abstract}
 It was shown recently that extended black hole thermodynamics, where the cosmological constant is a dynamical variable, giving rise to a  pressure $p$ and its conjugate volume $V$, can be given a natural setting in the context of braneworld models. We study the specific heat capacities $C_p(T)$ and $C_V(T)$ of the quantum version of the BTZ black hole that lives in the induced gravity theory on the brane. There are multiple branches of solutions, and we explore and characterize  key features of the possible behaviour. We identify and study  a critical point in the space of solutions where both specific heats diverge. In the regime of weak backreaction where we are close to an ordinary theory of gravity, the black hole is ``sub-entropic'', but as backreaction is increased we note that there are parts of parameter space that  has regions where it is ``super-entropic''. While a study of the sign of the specific heats does not always show a corresponding  instability (conjectured in the literature), the presence  of strong backreaction makes interpretation unclear.
\end{abstract}

\maketitle

\section{Introduction}
Quantum mechanics makes the semi-classical physics of black holes compatible with the laws of thermodynamics~\cite{Bekenstein:1973ur,Bekenstein:1974ax,Hawking:1975vcx,Hawking:1976de}, imparting a temperature $T{=}\kappa/2\pi$ and an entropy $S{=}A/4$ to them\footnote{We work in units where $G{=}k_B{=}c{=}\hbar{=}1$.}, where~$\kappa$ is the surface gravity of the event horizon and $A$ is the black hole horizon's surface area~\cite{Bardeen:1973gs}. The result is that a set of laws of black hole thermodynamics emerge, the first of which is $dM=TdS$, where~$M$ is the black hole's mass, playing the role of  internal energy,~$U$. 
It is interesting to calculate the specific heat capacity. Famously, in asymptotically flat space, the $4D$ Schwarzschild black hole has a temperature given by $T{=}1/8\pi M$, and an entropy $S{=}\pi r_{+}^2{=}4\pi M^2$, where $r_{+}{=}2M$ is the horizon's radius. This means that black hole has a \emph{negative} specific heat capacity,  $C(T){=}{-}1/8\pi T^2$, and hence is unstable~\cite{Hawking:1982dh}.

The situation changes when considering asymptotically anti-de Sitter (AdS) black hole solutions. Not only does the AdS spacetime act as a natural confining box, but there is also a negative cosmological constant, $\Lambda$, which plays the role of  a positive dynamical pressure given by $p{=}{-}\Lambda/8\pi$. Note that in~$D$ spacetime dimensions, $\Lambda$ defines a natural length scale of the AdS geometry {\it via} $\Lambda{=}{-}(D{-}1)(D{-}2)/2 L^2$. The presence of a  pressure~$p$ suggests (if it could be made dynamical--see footnote~\ref{footy}) an extension of  the first law of black hole thermodynamics to include a pressure-volume work term.\footnote{See refs.~\cite{Henneaux:1984ji,Teitelboim:1985dp,Henneaux:1989zc} for some early alternative approaches.}
%
An exploration of  this idea results~\cite{Kastor:2009wy} in an ``extended'' black hole thermodynamics where the mass no longer corresponds to the internal energy~$U$, but rather to the enthalpy, $M{=}H{\equiv} U{+}pV$. In other words the first law becomes $dM{=}TdS{+}Vdp$. The  temperature is hence $T{=}\left.{{\partial M}{/\partial S}}\right|_p$ and the thermodynamic volume now appears as a new quantity in the theory given by $V{=} \left.{{\partial M}/{\partial p}}\right|_S$. Note that in various simple  cases it does correspond to a physical spacetime volume, but it is a more general function of  parameters away from such circumstances~\cite{Cvetic:2010jb}, as will be seen in examples below.

With $p$ and $V$ present, one must specify which is being held constant when defining a specific heat, giving two quantities of interest, which may be written as  $C_p{=}dH/dT {=} d(U{+}pV)/dT$ and $C_V{=}dU/dT$, respectively.  The usual specific heat in ordinary black hole thermodynamics is the former. For example, for Schwarzschild-AdS black holes, there exist two families of solutions~\cite{Hawking:1982dh} corresponding to black holes that are large or small compared to the AdS length scale $L$. The small black holes have negative~$C_p$, just as their asymptotically flat counterparts do. The large black holes possess a \emph{positive}~$C_p$ and are thus stable (which makes them so useful for studying phases of the holographically dual thermal Yang-Mills theory~\cite{Witten:1998zw}). On the other hand, for both large and small Schwarzschild-AdS black holes, $C_V(T){=}0$, which follows~\cite{Dolan:2010ha} from the fact that for static black holes both the entropy and the volume are powers of the horizon radius $r_+$ and hence are not independent functions.

An industry sprung up as a result of this framework, based on studying various AdS black hole systems in extended thermodynamics (see ref.~\cite{Kubiznak:2016qmn} for a review) and  comparing the results (such as phase transitions  and associated critical exponents) to the thermodynamics of systems made from ordinary ({\it i.e.} non-gravitational) systems. While it is  an interesting program, a cautionary note pointed out in ref.~\cite{Johnson:2019vqf} is the fact that the quantity $C_V$ is an   important measure of {\it the available degrees of freedom of a  thermodynamic system}, at least as defined in traditional thermodynamics as those that can be given energy without the system doing work. Therefore, comparisons between systems based on critical exponents alone should be supplemented by the knowledge gained about the degrees of freedom from~$C_V$. In fact, many  of the simplest black hole systems in extended thermodynamics have {\it zero}   degrees of freedom by this measure (see the  above Schwarzschild-AdS case, as well as the Reissner-Nordtr\"{o}m-AdS case for which there is nonetheless a rich van der Waals-like phase structure\cite{Chamblin:1998pz,Kubiznak:2014zwa}), or only have relatively sparse such degrees of freedom, as shown by the characteristic presence~\cite{Johnson:2019vqf} of Schottky peaks in $C_V(T)$ after addition of  various kinds of charges, or rotation. 

An important $D{=}3$ asymptotically AdS geometry is the Ba\~{n}ados, Teitelboim, and Zanelli (BTZ) black hole~\cite{Banados:1992wn,Banados:1992gq}. For horizon radius $r_+$ its extended thermodynamic quantities are~\cite{Frassino:2015oca}: 
\begin{equation}
    \label{cBTZ}
    M=\frac{4pS^2}{\pi},\;\; T=\frac{8pS}{\pi},\;\; V=\frac{4S^2}{\pi}, \;\; 
    S=\frac{\pi}{2}r_{+}\ ,
\end{equation}
from which it follows that $C_p(T){=}\pi T/8p{>}0$ and $C_V(T){=}0$, somewhat in analogy with the large AdS-Schwarzschild black holes in higher dimensions.

It was noted in ref.~\cite{Johnson:2019mdp}  that in considering the stability of black holes in this extended framework, the sign of {\it both} $C_p$
 and  $C_V$ must be taken into account. Even with a positive $C_p$ a system can find itself with a negative $C_V$ in extended thermodynamics, thereby possessing an instability in that context.  A simple example of this arises when electric charge~$Q$ is included for the BTZ black hole. While $C_p$ remains positive,  the expression above for the entropy remains the same while the deformation by $Q$ reduces the volume, giving $V{=}{4S^2}/{\pi}{-}Q^2/32 p$. This gives a non-zero $C_V$ and in fact  $C_V(T){=} {-}Q^2/32T\times f(Q,p,T)$, where the function $f$ is positive~\cite{Johnson:2019mdp}. The solution is unstable in extended thermodynamics.

 It is worth trying to understand the origin and meaning of such instabilities. An important (but not the only) source was suggested in ref.~\cite{Johnson:2019mdp} to be the fact that the solution is ``super-entropic", which is to say that its entropy-to-volume ratio exceeds that which an AdS-Schwarzschild black hole of the same volume would have. Put differently, it is said to violate the ``reverse isoperimetric identity'', which means the following~\cite{Cvetic:2010jb}: In $D=3$, there is a quantity ${\cal R}$ that should satisfy: 
 \begin{equation}
 \label{eq:isoperimeter}
     {\cal R}\equiv\left[\frac{\pi V}{4S^2}\right]^\frac12 \geq 1\ .
 \end{equation}
 The connection to the instability comes because ref.~\cite{Johnson:2019mdp} derived the following  exact form for $C_V$:
 \begin{equation}
 \label{eq:superentropicCV}
     C_V= -S\left(\frac{4S^2-\pi V}{12 S^2 - \pi V}\right) = -S\left(\frac{1-{\cal R}^2}{3 - {\cal R}^2}\right)\ ,
 \end{equation} and so because ${\cal R}<1$ for that solution (``too much'' entropy) the instability is present. Given the examples available at the time, a natural conjecture~\cite{Johnson:2019mdp}  was that the sign of $C_V$ should be negative when ${\cal R}{<}1$, suggesting that the instability could be traced to some simple feature of how the system's entropy comes about, perhaps at the microscopic level. (Indeed, ref.\cite{Johnson:2019wcq}) succeeded in finding microscopic evidence (using a dual CFT$_2$ language) for such an instability in this and an even larger class of BTZ-based models.)

It is important to note that the conjecture does not suggest that all thermodynamic instabilities should be attributed to being super-entropic, but rather that being super-entropic should lead to an instability, evident in $C_V$ becoming negative. (This is consistent with the fact that another way of having~$C_V$ become negative in equation~(\ref{eq:superentropicCV}) is to have~${\cal R}^2{>}3$. This would mean that there is an instability, but it is not connected to being super-entropic.)

Subsequent work has shown that the situation is a bit  more complicated. Ref.~\cite{Cong:2019bud} argued that there is a family of generalized BTZ models that furnish super-entropic cases where ${\cal R}{<}1$ with $C_V$ positive, but ref.~\cite{Johnson:2019wcq} showed that a closer examination using a definition of super-entropicity that defines ${\cal R}$ in terms of entropy (instead of area as done there) does not yield the suggested counter examples.  Since then there have been some newer examples suggested  to be super-entropic while stability seems to be preserved~(see {\it e.g.,} refs.\cite{Song:2023zre,Jing:2020sdf,JahaniPoshteh:2021clv,Song:2023kvq,He:2023tiz}). However, ref.~\cite{Appels:2019vow} has cast doubt on whether the exotic types of solutions being used can really be classified as super-entropic. So it seems that the situation needs further study.

Therefore, finding new tractable examples of non-trivial behaviour for $C_p$ and $C_V$ could be of  value. In this paper we present computations of and observations about the specific heat capacities of the ``quantum'' version of the BTZ black hole (qBTZ) which  arises from computing  a backreaction in a braneworld model~\cite{Emparan:1999fd,Emparan:2020znc}. Ref.~\cite{Frassino:2022zaz} has recently presented the extended thermodynamics of this system\footnote{\label{footy} In fact, because the  cosmological constant (and hence pressure) in the AdS$_3$ arises from the tension of a Karch-Randall~\cite{Karch:2000ct} brane embedded in AdS$_4$,  their work shows that the extended thermodynamics has (at least) one framework  where it is entirely natural to allow them to be dynamical.}.

Much like in the Schwarzschild-AdS case, the qBTZ possesses different branches corresponding to different classes of behaviour of black hole solutions, and a quantum analogue of the Hawking-Page structure is evident. However, unlike Schwarzschild-AdS, we will observe that both $C_p(T)$ and $C_V(T)$ possess negative and positive branches in certain ranges of the temperature. While we have not exhaustively examined all regions of parameter space (due to the lack of closed-form heat capacities written explicitly as functions of $T$ and $V$), we explore some regions enough to establish some instructive features. Along the way we observe a curious critical point in the thermodynamics and study some of the physics in its neighborhood. The quantities  $C_p$ and $C_V$ diverge  at this point, showing  a phase transition. 

Crucially, at leading order in a weak backreaction expansion, where the theory of gravity on the brane is not too exotic, the quantity ${\cal R}$ is greater than one, and although the specific heats can have both signs in this regime, the conjecture of ref.~\cite{Johnson:2019mdp} is not in question. We observe that higher order corrections and certainly in the fully backreacted case, there seem to be regions where the system can be super-entropic while also having positive specific heats.  While this {\it could} mean that the conjecture fails for these examples, it is not really clear if the discussion really applies at all in this case, given that the induced gravity on the Karch-Randall brane is of a very different character from the starting context. 

While we were completing this work, we learned of forthcoming complementary work by Frassino, Pedraza, Svesko and  Visser~\cite{Frassino:2023wpc}, which carries out a careful study of the thermodynamic phase structure of the extended thermodynamics of this same qBTZ system. Although the foci of our works differ, some of our results overlap.

\section{The Quantum BTZ Black Hole}

\subsection{Review and observations}
\label{sec:review}
The qBTZ black hole arises naturally in the context of holographic braneworld models and   many of its thermodynamic quantities can be computed explicitly to all orders in the backreaction parameter. See refs.~\cite{Emparan:1999fd,Emparan:2020znc} for details. The setting is a Karch-Randall~\cite{Karch:2000ct} AdS$_3$ brane embedded in a sliced AdS$_4$ C-metric. There is an induced black hole in the AdS$_3$, which is the qBTZ black hole. 
The brane's position is parameterized by a parameter $\ell$, which is inversely proportional to both the tension of the brane and the acceleration of the black hole in the C-metric~\cite{Emparan:2020znc,Frassino:2022zaz}. The brane's tension and the AdS$_4$ cosmological constant in turn induce a cosmological constant on the AdS$_3$, which in turn sets the pressure $p$. In this paper we will choose, following ref.~\cite{Frassino:2022zaz}, to hold the length scale, $L_4$, of the AdS$_4$ bulk geometry fixed and set to unity. This results in  $dp=-d\tau$ where $\tau$ is the tension in the brane, with no further contributions to $dp$ from variations in $L_4$. Hence,  variations in the $D{=}3$ pressure are related to variations in the brane tension with no  contributions from other parameters\footnote{Assuming fixed Newton's constants $G_3$ and $G_4$, of course.}.

Appendix~\ref{appendixMTS} gives the full details of the derivation of $M, T$, and $S$, which are
\begin{eqnarray}
    \label{qM}
    M(\nu,z)&=&\frac{\sqrt{1+\nu^2}}{2G_3} \frac{z^2(1-\nu z^3) (1+\nu z)}{(1+3z^2+2\nu z^3)^2}\ ,
\\
    \label{qT}
    T(\nu,z)&=&\frac{1}{2\pi \ell_3(\nu)} \frac{z(2+3\nu z+\nu z^3)}{1+3z^2+2\nu z^3}\ ,\,\,{\rm and}\quad
\\
    \label{qS}
     S(\nu,z)&=&\frac{\pi \ell_3(\nu) \sqrt{1+\nu^2}}{G_3}\frac{z}{1+3z^2+2\nu z^3}\ ,
\end{eqnarray}
where we have temporarily restored the factors of the $D{=}3$ Newton constant, $G_3$, for clarity\footnote{The chosen value of fixed $L_4$ and $G_4$ determines $G_3$ via $G_3{=}G_4/2L_4$. Having fixed $L_4{=}1$, we will fix $G_4{=}2$ for the sake of having $G_3{=}1$ in the remainder of this paper.}.
The two dimensionless parameters, $\nu$ and $z$, are defined as 
\begin{equation}
\label{rrelation}
    \nu\equiv \frac{\ell}{\ell_3}\ , \quad {\rm and} \quad z\equiv\frac{\ell_3}{x_1 r_{+}}\ ,
\end{equation} 
where $\ell_3$ is the ``bare" AdS$_3$ length scale, without taking any of the backreaction effects into account, and appears in both $L_4$ and $L_3$, the latter being the AdS$_3$ length scale with backreaction effects included (see below). The parameter  $x_1$ is the single positive root remaining in a polynomial present in the C-metric after part of the bulk is cut off by the brane, and $r_{+}$ is the location of the event horizon. Both $\nu$ and $z$  take values in the interval $[0,\infty)$, where small values of $\nu$ correspond to small backreaction. Furthermore, equations~(\ref{qM}), (\ref{qT}), and (\ref{qS}) satisfy $dM{=}TdS$, if $\nu$ is held fixed and only $z$ is allowed to vary. This directly implies the independence of the pressure $p$ on $z$. Note that we have written $\ell_3=\ell_3(\nu)$ in preparation for the discussion to follow.

\begin{figure*}
     \centering
     \begin{subfigure}[h]{0.32\textwidth}
         \centering
         \includegraphics[width=\textwidth]{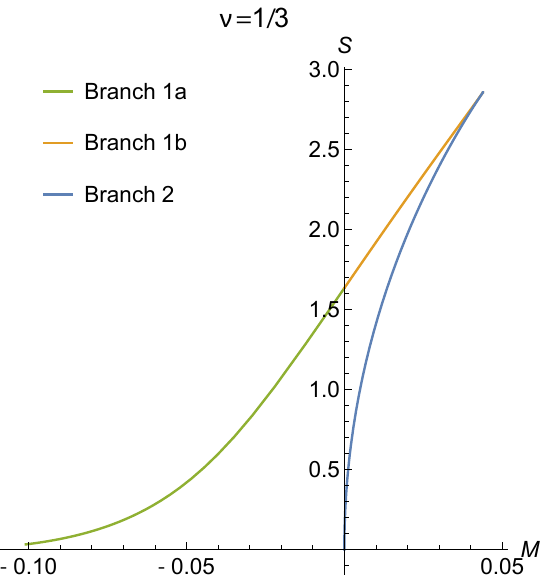}
         \caption{$S$ {\it vs.} $M$ for $\nu=1/3$.}
         \label{fig:SvsMexact}
     \end{subfigure}
     \hfill
     \begin{subfigure}[h]{0.32\textwidth}
         \centering
         \includegraphics[width=\textwidth]{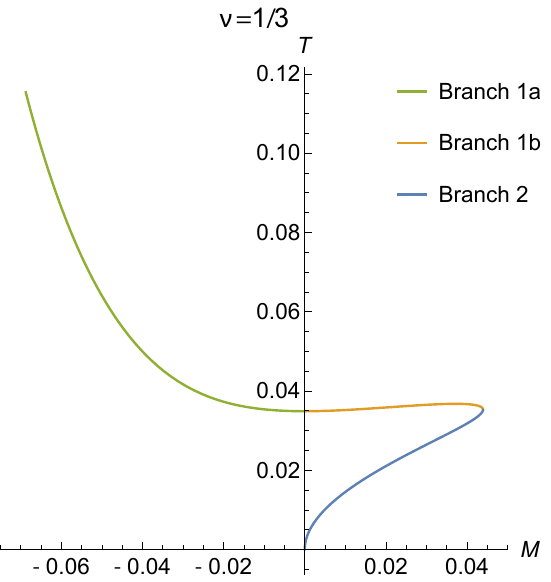}
         \caption{$T$ {\it vs.} $M$ for $\nu=1/3$.}
         \label{fig:TvsMexact}
     \end{subfigure}
     \hfill
     \begin{subfigure}[h]{0.33\textwidth}
         \centering
         \includegraphics[width=\textwidth]{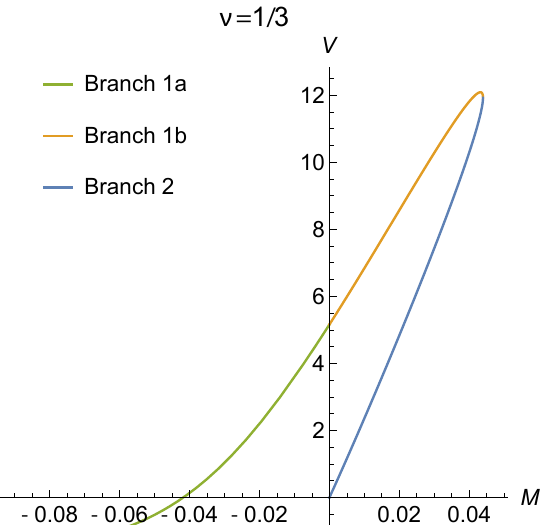}
         \caption{$V$ {\it vs.} $M$ for $\nu=1/3$.}
         \label{fig:VvsMexact}
     \end{subfigure}
        \caption{The three branches corresponding to the three possible black holes at fixed $\nu{=}1/3$. For positive $M$, there exist two branches, the lower one corresponding to stable black holes with positive heat capacity and the upper one corresponding to unstable black holes with negative heat capacity. The heat capacity for the $M{<}0$ branch is also negative. Note that in these plots, $z$ is an implicit parameter which increases as the plots are traced starting with Branch 2 and ending with Branch 1a.}
        \label{fig:STVvsMexact}
\end{figure*}

The quantities $\ell, \ell_3, L_4$ and $L_3$ satisfy

\begin{equation}
    \label{L4}
    \frac{1}{L_4^2}=\frac{1}{\ell^2}+\frac{1}{\ell_3^2}\ ,\quad {\rm and} \quad
    \frac{1}{L_3^2}=\frac{2}{L_4^2}\bigg(1-\frac{L_4}{\ell}\bigg)\ .
\end{equation}
Note from the first relation in equations~(\ref{L4}) that fixing $L_4$ imposes a functional dependence between $\ell$ and $\ell_3$, which can be re-expressed via a dependence of $\ell_3$ on $\nu$ given by $\ell_3(\nu)=\sqrt{1+\nu^2}/\nu$. The pressure on the brane is given by $p=1/8\pi L_3^2$, which is given in terms of $\nu$ by

\begin{equation}
    \label{qP}
    p(\nu)=\frac{1+\nu^2-\sqrt{1+\nu^2}}{4\pi \ell_3^2 (\nu) \nu^2}.
\end{equation}

Having the pressure and $\ell_3(\nu)$, one can then seek a thermodynamic volume to complete the First Law, $dM=TdS+Vdp$. One can construct this volume directly from the First Law, it is given by

\begin{equation}
    \label{generalqV}
    V=\left({\left.\frac{\partial p}{\partial \nu}\right|_z}\right)^{-1}\left[\left.\frac{\partial M}{\partial \nu}\right|_z-T\left.\frac{\partial S}{\partial \nu}\right|_z\right]\ .
\end{equation}

\noindent Explicitly calculating (\ref{generalqV}) yields~\cite{Frassino:2022zaz}:

\begin{widetext}
\begin{equation}
    \label{qV}
    V(\nu,z)=\frac{-2\pi \ell_3^2 (v) z^2(-2+\nu^2+3\nu^3 z^3+\nu z(\nu^2-4)+\nu^4 z^4)}{(1+3z^2+2\nu z^3)^2}.
\end{equation}
\end{widetext}

We will work with the preceding thermodynamic functions to all orders of $\nu$ and $z$. The first surprising feature is that for a fixed, finite $\nu$, the allowable masses exist on the finite interval

\begin{equation}
    \label{massinterval}
    -\frac{\sqrt{1+\nu^2}}{8}\leq M\leq\frac{\sqrt{1+\nu^2}}{24}.
\end{equation}

This result was previously obtained in refs. \cite{Emparan:2002px,Emparan:1999fd}. We will restrict ourselves in this paper to working in the regime of nonnegative $M, T, S,$ and $V$, though we present plots in figure \ref{fig:STVvsMexact} that include the negative mass and negative volume branches of these quantities for completeness. Plotting $S$, $T$, and $V$ {\it vs.}~$M$ for fixed, finite $\nu$ reveals three distinct branches, two of which exist for positive $T, S, V$ and~$M$, and one of which exists for negative $M$ but positive $T, S$ and~$V$ (as $z$ increases, $V$ eventually becomes negative as well).

Following ref. \cite{Emparan:2020znc}, we label the three branches~$1a$, $1b$, and $2$. In each case, Branch $2$ originates at the origin and rises in the first quadrant as $M$ approaches its upper bound. There, it joins Branch~$1b$, and in the case of the $S$ {\it vs.} $M$ plot, the two branches join at a cusp, very reminiscent of the joining of branches for Schwarzschild-AdS~\cite{Hawking:1982dh}. Branch $1b$ originates on the vertical axis, for positive $T$, $S$, or~$V$ (and for $M{=}0$), before it joins Branch 2. Branch~$1a$ exists in the second quadrant as the only solution there, and it joins branch $1b$ at $M=0$. See figure~\ref{fig:STVvsMexact}.

The cusp present in the $S$ {\it vs.}~$M$ plot at constant $\nu$ is worth considering further. Its location  corresponds to where $S(z)$ attains its  maximum value as a function $z$ at fixed $\nu$. Below this point, the entropy function is split into two distinct branches. Note that at fixed~$\nu$, $\ell_3$ is also fixed. Further, $x_1$ must also be fixed because it is a root of a polynomial that is independent of $\ell$ and $\ell_3$. Therefore, at fixed $\nu$, $z$ depends solely on $r_+$ and goes like $1/r_+$ (see equation~(\ref{rrelation})). Thus, for a given horizontal slice, the corresponding black holes on the  branches represent two different masses that can be attributed to a large (Branch 2) and small (Branch 1b) black hole of equal entropy.

Between $z{=}0$ and the location of the entropy's cusp, {\it i.e.,} on Branch 2,  the black hole  shrinks down from infinite size, while its entropy and temperature rise. It has positive specific heat. Meanwhile on Branch 1b, moving down from the cusp  $z$ continues to increase, so there is further shrinking of the black hole. Here the entropy  entropy  falls. For some values of $\nu$ the temperature continues to rise and thus the specific heat capacity must be negative, while for other values of $\nu$ the temperature initially begins to drop, but it will eventually reverse course and continue rising. The analysis is made more complex by the existence of three temperature branches while only two entropy branches exist. For a range of fixed values of $\nu$ the black hole's temperature will begin to fall beyond the cusp, but this fall is only temporary. Eventually, once $z$ rises enough, the temperature will begin to rise again as the black hole continues to shrink. As will be clear later, it is only at a critical point that this intermediate falling temperature branch is absent. 

Note that an apparent contradiction seems present:  For small~$z$, the black hole must be very large, and yet the $V$ {\it vs.} $M$ plot reveals that at small $z$, $V$ is very small. This is resolved by recalling the discussion in the Introduction.  The geometric volume built from $r_+$ is distinct from the thermodynamic volume $V$~\cite{Cvetic:2010jb}. Indeed in the  Introduction the electrically charged BTZ case was briefly discussed and there the $V$ was seen to be reduced by increasing the charge $Q$ although the geometric volume remained unchanged. 

\begin{figure*}
     \centering
     \begin{subfigure}[h]{0.32\textwidth}
         \centering
         \includegraphics[width=\textwidth]{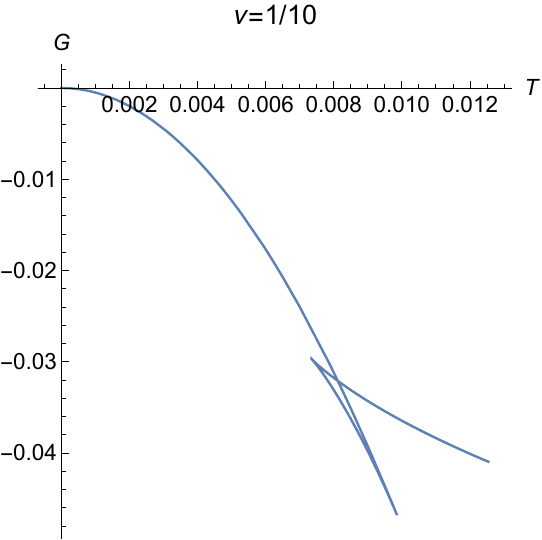}
         \caption{$G$ vs. $T$ for $\nu=\nu_c/10$.}
         \label{fig:Gbelowcrit}
     \end{subfigure}
     \hfill
     \begin{subfigure}[h]{0.32\textwidth}
         \centering
         \includegraphics[width=\textwidth]{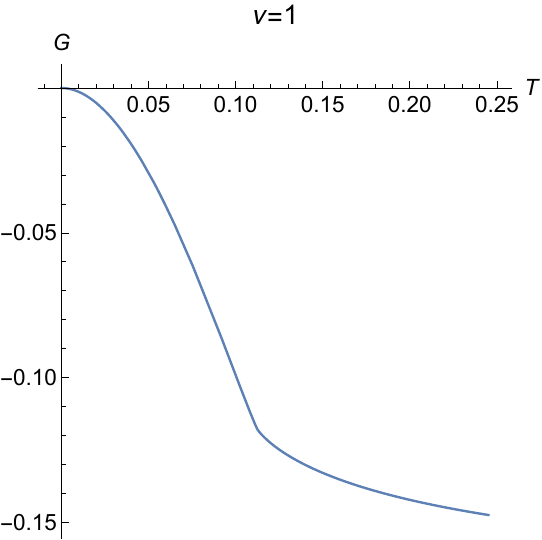}
         \caption{$G$ vs. $T$ for $\nu=\nu_c$.}
         \label{fig:Gcrit}
     \end{subfigure}
     \hfill
     \begin{subfigure}[h]{0.32\textwidth}
         \centering
         \includegraphics[width=\textwidth]{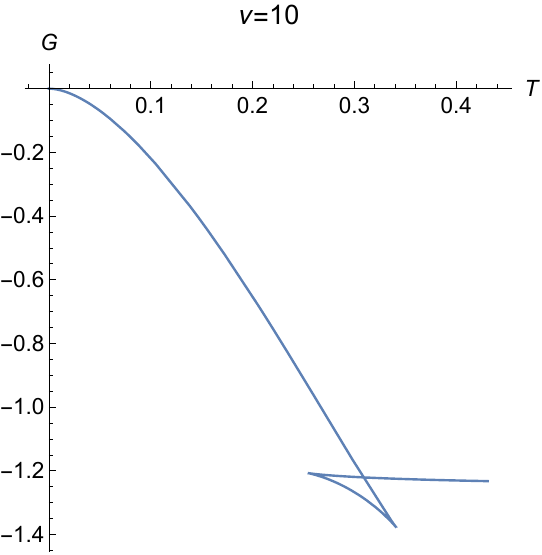}
         \caption{$G$ vs. $T$ for $\nu=10\nu_c$.}
         \label{fig:Gabovecrit}
     \end{subfigure}
        \caption{The Gibbs free energy as a function of temperature with pressure fixed to values below, at, and above the critical pressure. The characteristic swallowtail pattern emerges for \emph{both} pressures below as well as pressures above the critical pressure. At the critical pressure the plot displays a kink, as expected.}
        \label{fig:Gibbs}
\end{figure*}

\subsection{Specific heat at constant pressure}
With the  thermodynamic quantities in hand, it is possible to derive specific heat capacities for the qBTZ as functions of $\nu$ and $z$.  We begin with $C_p$, noting that constant $p$ is equivalent to fixed $\nu$, yielding\footnote{After a relabelling of parameters, it can be seen that this result was obtained already in ref.~\cite{Kudoh:2004ub}. We thank Andrew Svesko for pointing this out.}:
\begin{widetext}
\begin{equation}
    \label{Cpvz}
    C_p(\nu,z)=T(\nu,z) \frac{\partial S}{\partial z}\bigg\rvert_\nu \frac{\partial z}{\partial T}\bigg\rvert_\nu=\frac{\pi z \sqrt{1+\nu^2}\ell_3(\nu) (3z^2-1+4\nu z^3)(2+3\nu z+\nu z^3)}{2(1-\nu z^3)(1+3z^2+2\nu z^3)(3z^2-1-3\nu z+\nu z^3)}.
\end{equation}
\end{widetext}
Since pressure is fixed by fixing $\nu$, this expression for~$C_p$ is relatively easy to analyze at face value. (This will not be the case for $C_V$ since  fixed~$V$  is a nontrivial curve in the $(\nu,z)$ plane--see later.) {Note that in calculating {(\ref{Cpvz}}) we  used the definiton of heat capacity given by $C{=}T\partial S/\partial T$, and employed the chain rule when evaluating the left side of {(\ref{Cpvz}}), exploiting the fact that $T$ becomes single-valued when~$\nu$ is fixed. }

Notice a potential divergence in $C_p$ coming from 
$1-\nu z^3=0$ and $3z^2-1-3\nu z+\nu z^3=0$. These two curves intersect at $(\nu,z)=(1,1)$, and it is  tempting to conclude that this is an interesting  point. It is, and we will soon prove that this is actually a critical point. 
Generically,   a critical point simultaneously solves the equations:

\begin{equation}
    \label{criticalpt1}
    \frac{\partial p}{\partial V}\bigg\rvert_T=0\;\;\;\;\text{and}\;\;\;\;\frac{\partial^2 p}{\partial V^2}\bigg\rvert_T=0.
\end{equation}

Performing this differentiation exactly is very difficult, due simply to the highly implicit nature of the equations defining our thermodynamic quantities in this case. In the previous section we noted during our investigation of the branches that unlike the entropy, which possesses a lone local maximum as a function of $z$, the temperature contains a local minimum as well as a local maximum for finite $z>0$ for certain ranges of the value of $\nu$. The local maximum, and then minimum, arise in Branch $1b$ as~$z$ increases. From there, as~$z$ continues to grow,~$T$ grows without bound as well. Since the temperature continues to rise even after the entropy has begun to fall, we concluded that this corresponds to an unstable branch, whereas both the temperature and entropy rise in Branch 2 as~$z$ grows, making that the stable branch. Observing the temperature for a range of  values of~$\nu$ shows that as $\nu$ rises from zero to one, the local maximum and minimum move towards each other. At $\nu{=}1$, they merge into an inflection point corresponding to~$z{=}1$. At higher values of $\nu$, they separate again into disparate extrema. It turns out that in addition to the condition imposed on the first and second derivatives of the temperature with respect to $z$, there is also a condition that the derivative of the entropy with respect to $z$ must satisfy. The details are worked out in Appendix \ref{appendixcritical},  with result:
\begin{equation}
    \label{criticalpt2}
    \frac{\partial T}{\partial z}\bigg\rvert_\nu=0,\;\;\;\;\frac{\partial^2 T}{\partial z^2}\bigg\rvert_\nu=0,\;\;\;\;\text{and}\;\;\;\;\frac{\partial S}{\partial z}\bigg\rvert_\nu \neq 0.
\end{equation}

The fact that this maximum and minimum begin as separate entities, merge into a stationary point, and then separate again results in novel characteristics of these black holes. In other black hole systems,  after the critical point the system enters a new phase and usually remains there, with no further divergences or multi-valued branches appearing in the heat capacity, 
but for these qBTZ systems there is a similarity  between small and large $\nu$. That is to say, for both $\nu$ sufficiently below $\nu_c$ \emph{and} sufficiently above~$\nu_c$, the heat capacity possesses several branches and multivaluedness.

A useful quantity that supports these observations about  the critical point and its neighborhood is the Gibbs free energy $G{=}H{-}TS$. Plotting  $G(T)$ at fixed~$p$ demonstrates the characteristic swallowtail pattern below the critical pressure. At the critical pressure the Gibbs free energy becomes single-valued, but there is a jump discontinuity in its derivative at the critical temperature. Above the critical pressure the swallowtail pattern reemerges, whereas we might have expected the function  to remain smooth as happens for  other systems \cite{Kubiznak:2012wp,Altamirano:2013ane,Altamirano:2013uqa}. See figure~\ref{fig:Gibbs}. To our knowledge, this is not observed in other black hole systems studied in the literature.

\begin{figure}[h]
         \centering
         \includegraphics[width=0.43\textwidth]{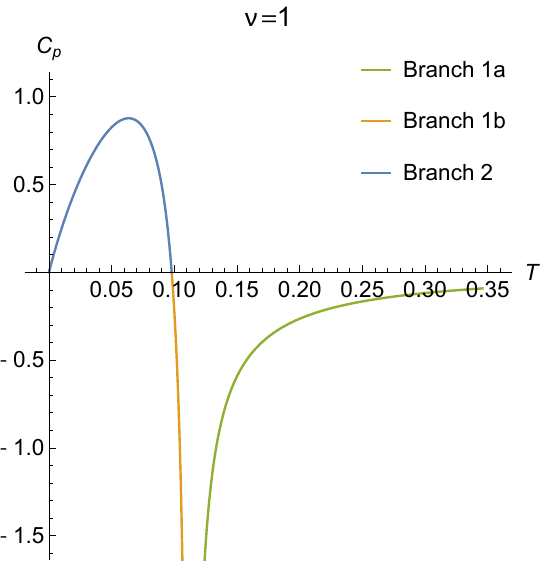}
         \caption{$C_p(T)$ for $\nu=\nu_c$.}
         \label{fig:Cnuatnucrit}
     \end{figure}

The \emph{unique} point satisfying (\ref{criticalpt2}) is $(\nu_c, z_c){=}(1,1)$. Furthermore, when plotting $C_p(T)$, it is these values of $\nu$ and $z$ which produce a very strong divergence, the tell-tale sign of a first-order phase transition.  See figure~\ref{fig:Cnuatnucrit}. 
     \begin{figure}[h]
         \centering
         \includegraphics[width=0.43\textwidth]{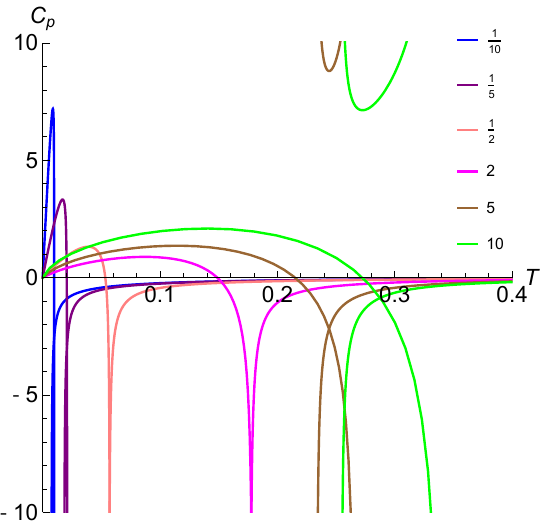}
         \caption{$C_p(T)$ for various $\nu$ in the neighborhood of~$\nu_c$.  Different values of $\nu/\nu_c$ are shown.}
         \label{fig:Cnuatvariousnu}
     \end{figure}
The critical values of $\nu$ and $z$ yield the following critical values for the thermodynamic functions:
\begin{align}
        M_c&=0&T_c&=\frac{1}{2\sqrt{2}\pi}&p_c&=\frac{2-\sqrt{2}}{8\pi}\\
        S_c&=\frac{\pi}{3} & V_c&=0 \nonumber
             \label{criticalpoint}
\end{align}


  

Based on these values, we conclude that this is the  transition between Branch $1b$ and~Branch~$1a$. It is a first order phase transition, following from the fact that there is  both a jump in the entropy as well as a kink in the Gibbs free energy. Figure~\ref{fig:Cnuatvariousnu} shows sample behaviour of $C_p(T)$ in the neighborhood of the critical point. At the critical pressure, the function is single-valued and displays a strong divergence at $T{=}T_c$. Away from the critical pressure, both below and above it, the function becomes multivalued and displays multiple branches.

\subsection{Specific heat at constant volume}
\begin{figure}[t]
     \centering
         \includegraphics[width=0.47\textwidth]{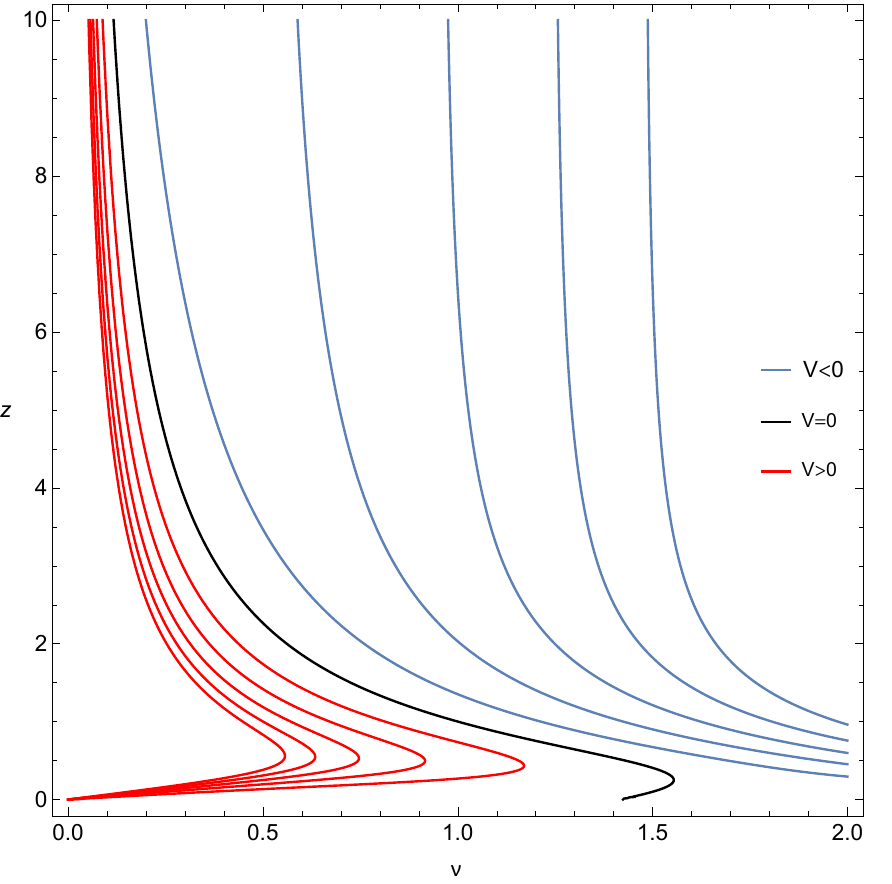}
        \caption{Sample isochors in the  $(\nu,z)$ plane. Red curves denote positive volumes, from $V=5$ (left) to $V=1$  (right). The black curve is  $V=0$. The blue curves  range from $V=-1$ to $V=-5$. 
        }
        \label{fig:isochors}
    \end{figure}

Consider the family of isochors in the $(\nu,z)$ plane, as shown in figure~\ref{fig:isochors}. At fixed $V{\geq}0$ there are two distinct branches of an isochor,  corresponding to two families of black holes, analogous to the fixed~$p$ ({\it i.e.,}~$\nu$) case.

We state here a general equation for $C_V(\nu,z)$ in terms of partial derivatives of the thermodynamic functions, and leave the details of the derivation to Appendix \ref{appendixCV}. We find that\footnote{We thank Seyed Mansoor for pointing out an error in the derivation of $C_V(\nu,z)$ in an earlier version of this paper.}
\begin{widetext}
\begin{equation}
    \begin{split}
        \label{CVvz}
    C_V(\nu,z)=T(\nu,z) \bigg[\frac{\partial S}{\partial z}\bigg\rvert_\nu-\frac{\partial S}{\partial \nu}\bigg\rvert_z \frac{\partial V}{\partial z}\bigg\rvert_\nu\bigg(\frac{\partial V}{\partial \nu}\bigg\rvert_z\bigg)^{-1}\bigg] \bigg[\frac{\partial T}{\partial z}\bigg\rvert_\nu-\frac{\partial T}{\partial \nu}\bigg\rvert_z \frac{\partial V}{\partial z}\bigg\rvert_\nu\bigg(\frac{\partial V}{\partial \nu}\bigg\rvert_z\bigg)^{-1}\bigg]^{-1}.
    \end{split}
\end{equation}
\end{widetext}

\noindent Evaluating (\ref{CVvz}) explicitly yields\footnote{Juan Pedraza  calculated this expression for $C_V(\nu,z)$ before us, and we thank him for sharing it with us.}

\begin{equation}
    \label{CVvz2}
    C_V(\nu,z)=\frac{\mathcal{N}(\nu,z)}{\mathcal{D}(\nu,z)},
\end{equation}

\begin{widetext}

\noindent where
    \begin{equation}
        \label{CVnumerator}
        \mathcal{N}(\nu,z)=-2\pi \ell_3 (\nu) \nu \sqrt{1+\nu^2} z (1+\nu z)^3 (2+3 \nu z+\nu z^3) (4z(1+\nu^3 z^3)+3\nu(\nu z-1- z^2+\nu z^3))
    \end{equation}

\noindent and
\begin{align}
         \label{CVdenominator}
    \mathcal{D}(\nu,z)=&(1+3z^2+2\nu z^3)(16-48z^2+4\nu^8 z^8(z^2-3)+8\nu^7 z^7(3z^2-5)-4\nu^3 z(3-4z^2+45z^4)\\& \nonumber +80\nu z(1-2z^2)-4\nu^2 (1-27z^2+56z^4)+4\nu^5 z (3+2z^2-15z^4+2z^6)-4\nu^4(3z^2-1\\& \nonumber +17z^4+21z^6)+3\nu^6 z^2(3+5z^2-15z^4+15z^6)).
\end{align}
\end{widetext}

\begin{figure*}
     \centering
     \begin{subfigure}[h]{0.44\textwidth}
         \centering
         \includegraphics[width=\textwidth]{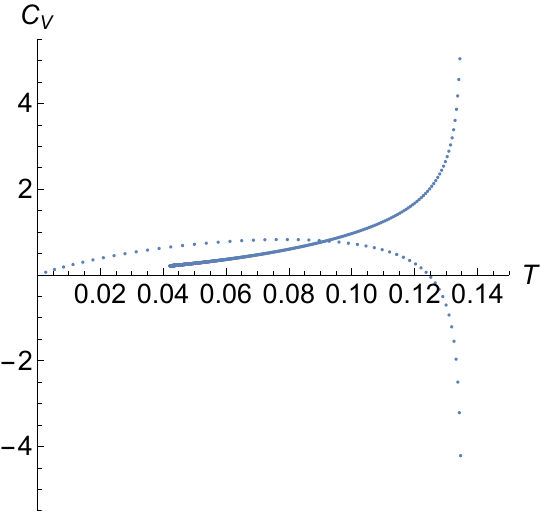}
         \caption{$C_V$ vs. $T$ for $V=V_c$.}
         \label{fig:CVatVcrit}
     \end{subfigure}
     \hfill
     \begin{subfigure}[h]{0.44\textwidth}
         \centering
         \includegraphics[width=\textwidth]{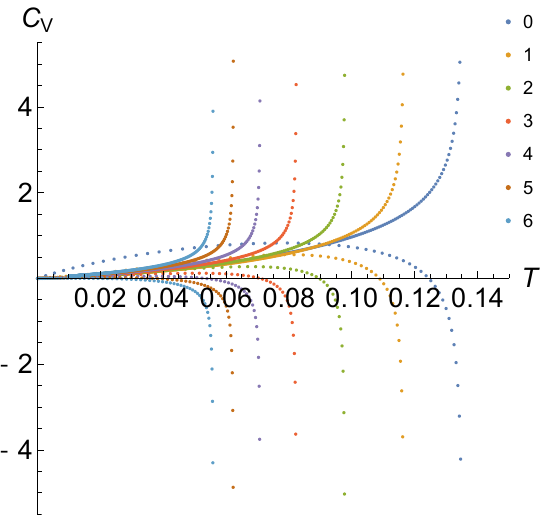}
         \caption{$C_V$ vs. $T$ for $V\geq V_c$.}
         \label{fig:CVatvariousV}
     \end{subfigure}
        \caption{$C_V(T)$ displays two branches for each fixed volume. Both begin at positive values of $C_V$, but as $T$ increases, one branch diverges to positive infinity, while the other one to negative infinity.The upper branch does not reach absolute zero, whereas the lower one does. (The legend on the right  shows the value of $V$.)}
        \label{fig:CVatVcritandothers}
\end{figure*}

Crucially, to interpret this expression correctly, we must still restrict ourselves to values of $\nu$ and~$z$ which lie on an isochor. This entails fixing $V$ in equation~(\ref{qV}) and then solving for, say,~$\nu$, resulting in a curve $\nu=\nu (z)$. Unfortunately, for arbitrary fixed $V$, the resulting polynomial cannot usually be solved analytically as it is of too high degree. Numerical explorations were employed for such cases. However, in the case of the critical isochor, $V_c{=}0$, the situation simplifies somewhat, since the coordinates lying on that isochor satisfy the relation $-2+\nu^2+3\nu^3 z^3+\nu z(\nu^2-4)+\nu^4 z^4=0$, which can be solved for $\nu=\nu(z)$. The result is not particularly illuminating, and thus we will omit it, however, its analysis allows us to exactly calculate an important feature of $C_{V_c}(T)$. Note that if one computes $T(\nu (z),z)$, {\it i.e.,} the temperatures at fixed $V=0$, and then takes the limit as~$z$ tends to infinity, the result is a finite temperature. That is to say, restricting ourselves to the critical isochor, while one of the branches of $C_V$ starts at $T=0$, the second branch terminates (as $z$ increases) at some $T_{\rm min}>0$. The exact result is very bulky, so we state it in Appendix \ref{appendixTmin}, but we can approximate it as $T_{\rm min}\simeq{0.0408}$. This pattern holds in general for different fixed $V$. Further, at each fixed $V$, the two branches of $C_V(T)$ diverge at some $T_{\rm div}$. Calculating this exactly is very difficult, but we can approximate ir in the case of $V_c$ as $T_{\rm div}\simeq{0.1355}$. See figure \ref{fig:CVatVcritandothers}.

A careful observation of isochors and isotherms in the $(\nu,z)$ phase space reveals that the two branches of the isochoric curves don't map exclusively onto the two  branches of  $C_V(T)$ (see figure~\ref{fig:CVatVcritandothers}). The reason for this is that the isothermal curves can sometimes intersect a given isochor branch twice. This complicates the analysis, and we leave a more in-depth exploration of the consequences of this feature to future work.

\section{Whither super-entropicity?}

It is possible to calculate  $\cal R$ of equation~(\ref{eq:isoperimeter}) to all orders of $\nu$ and $z$ and survey the regions where ${\cal R}{>}1$ and ${\cal R}{<}1$.  For the super-entropic (${\cal R}{<}1$) regions, the sign of $C_V$ can be examined to see if it is negative.  ($C_p$ can be examined too.)  A clear pattern did not emerge, but it must be immediately noted that it is not clear if  this teaches us {\it anything at all} in the regions where back-reaction is strong. 
For a start, strong backreaction means that the theory of gravity in question is no longer massless, and so it isn't {\it a priori} clear what rules should apply. 

A safer approach is to work in the limit of small backreaction parameter $\nu$, and it can be checked that this yields:
\begin{equation}
{\cal R}\simeq 1+z\nu-\frac14 (3+2z^2)\nu^2+O(\nu^3)\ ,
\end{equation}
The linear correction was already computed in ref.~\cite{Frassino:2022zaz}.
At  this order the black hole is  strictly sub-entropic, and although there are regions where $C_V(T)$ can go negative in this regime, it is not compelled to in order to confirm the conjecture of ref.~\cite{Johnson:2019mdp}. The next order may already be departing too far from ordinary gravity, but it is interesting  that it has a minus sign, and so can bring the black hole to super-entropicity. Examining such regions at this order shows that $C_V$ can remain positive in some of this super-entropic region, suggesting that the conjecture does not hold.  However, it must be stressed again that it is not clear at all if the conjecture should apply in its current form when there is significant backreaction, which takes the gravity theory induced on the Karch-Randall brane into unfamiliar territory.

 \begin{acknowledgments}
CVJ thanks Andrew Svesko for helpful comments and a delightful January conversation that renewed his interest in this area. Roberto Nazario thanks Juan Pedraza for sharing his expression for the heat capacity at constant volume. This work was supported in part by the  US Department of Energy  under grant  DE-SC 0011687.  CVJ thanks Amelia for her support and patience.    
\end{acknowledgments}

\appendix
\section{{Calculating $M$, $T$, and $S$}} 

\label{appendixMTS}
{We provide here abridged calculations for the expressions for $M$, $T$, and $S$ used in section~\ref{sec:review}. The full details of the calculations and discussions on the subtleties which arise are found in ref. {\cite{Emparan:2020znc}}. Recall that the setting described in the paper is an AdS$_4$ C-metric with an AdS$_3$ brane at the location of the regulator surface. The AdS$_4$ C-metric takes the form}
\begin{widetext}
\begin{equation}
    \label{C-metric}
    ds^2=\frac{\ell^2}{(\ell+xr)^2}(-H(r)dt^2+H(r)^{-1}dr^2+r^2(G(x)^{-1}dx^2+G(x)d\phi^2)),
\end{equation}
\end{widetext}
where
\begin{equation}
    \label{Hfxn}
    H(r)=\frac{r^2}{\ell_3^2}+\varkappa-\frac{\mu\ell}{r}
\end{equation}
and
\begin{equation}
    \label{Gfxn}
    G(x)=1-\varkappa x^2-\mu x^3.
\end{equation}
Note that in the preceding equations, $\mu{=}0$ corresponds to pure AdS and $\mu{>}0$ accounts for the holographic corrections to the quantum black hole. Furthermore, $\varkappa$ is a discrete variable that can take on the values -1, 0, and +1, with each value corresponding to a different induced geometry on the brane. In order to induce the BTZ geometry, one sets the value to -1, though we keep it here explicitly as $\varkappa$ for generality.

The metric induced on the brane by the backreaction (the brane is fixed at $x{=}0$) takes the form
\begin{equation}
    \label{induced metric}
    ds^2=-H(r)dt^2+H(r)^{-1}dr^2+r^2 d\phi^2\ .
\end{equation}
Note that in {(\ref{induced metric})} $\phi$ is $2\pi\Delta$ periodic, where
\begin{equation}
    \label{deltaterm}
    \Delta\equiv\frac{2x_1}{3-\varkappa x_1^2} .
\end{equation}
We normalize the coordinates to make $\phi$ $2\pi$ periodic by rescaling $t{=}\Delta \Bar{t}$, $\phi{=}\Delta \Bar{\phi}$, and $r{=}\Bar{r}/\Delta.$

We find $M$ in the usual manner by identifying the subleading constant term in $g_{\Bar{t}\Bar{t}}$ with $8{\cal{G}}_3 M$, where ${\cal{G}}_3{\equiv}G_3/\sqrt{1+\nu^2}$ is the effective Newton constant with backreaction effects taken into account. The result is
\begin{equation}
    \label{mass}
    M=-\frac{\varkappa\ell\Delta^2}{8G_3}=-\frac{\ell \varkappa x_1^2}{2G_3 (3-\varkappa x_1^2)^2} .
\end{equation}
Using the result from {\cite{Emparan:2020znc}} that
\begin{equation}
    \label{rxrelation}
    -\varkappa x_1^2=\frac{1-\nu z^3}{z^2(1+\nu z)} ,
\end{equation}
the result {(\ref{qM})} follows directly.
To find $T$, we may proceed as ref. {\cite{Emparan:2020znc}} did in the usual manner and perform a Wick rotation on the normalized metric and then impose periodicity on the imaginary time coordinate. The result is that
\begin{equation}
    \label{temp}
    T=\frac{\Delta H'(r_{+})}{4\pi} ,
\end{equation}
where
\begin{equation}
    \label{rplus}
    r_{+}^2=-\ell_3^2\varkappa \frac{ 1+\nu z}{1-\nu z^3}.
\end{equation}
Equivalently, we may simply employ {(\ref{qM})} directly and calculate $T$ \emph{via}
\begin{equation}
    \label{temp2}
    T\equiv \frac{\partial M}{\partial S}\bigg\rvert_p=\frac{\partial M}{\partial z}\bigg\rvert_\nu\left(\frac{\partial S}{\partial z}\bigg\rvert_\nu\right)^{-1}\ .
\end{equation}
The result {(\ref{qT})} follows directly from {(\ref{temp2})}.

Finally, we calculate $S$ using the Bekenstein-Hawking area law, calculating the area in the $D{=}4$ geometry and translating into $D{=}3$ quantities. This entropy is interpreted as the quantum-corrected, generalized entropy (though the generalized entropy and the standard entropy equal one another, see refs. \cite{Emparan:2020znc,Frassino:2022zaz,Emparan:2006ni} for details). The entropy is thus given by
\begin{equation}
    \label{entropy}
    S=\frac{1}{4G_3}\int_{0}^{2\pi\Delta} \,d\phi\int_{0}^{x_1}\,dx\\\frac{r_{+}^2 \ell^2}{(\ell+r_{+} x)^2}\ .
\end{equation}
Explicitly evaluating {(\ref{entropy})} while employing (\ref{rxrelation}) yields {(\ref{qS})}.

\section{Calculating $C_V(\nu,z)$}
\label{appendixCV}
We derive here a general formula for $C_V(\nu,z)$ for the qBTZ in terms of the thermodynamic quantities and their derivatives. We construct the heat capacity out of a product of $T$ with the ratio $dS/dT$, where $dS$ and $dT$ are the variations of $S$ and $T$, respectively. Recall that  for a function $f=f(x^1,x^2,...,x^N)$ of $N$ independent variables, the variation of $f$ is given by
\begin{equation}
    \label{variations}
    df=\frac{\partial f}{\partial x^i}dx^i,
\end{equation}
where the Einstein summation convention is being employed in (\ref{variations}). To impose the constant volume condition, we calculate the variation of $V$ and set $dV=0$. The result is a constraint between variations of $\nu$ and $z$, given by
\begin{equation}
    \label{constraintdV}
    d\nu=-\bigg(\frac{\partial V}{\partial \nu}\bigg\rvert_z\bigg)^{-1} \frac{\partial V}{\partial z}\bigg\rvert_\nu dz.
\end{equation}
\\
Imposing (\ref{constraintdV}) on $C{=}TdS/dT$ gives the result in (\ref{CVvz}). There is a benefit to writing $C_V(\nu,z)$ in this form. Typically, one has a choice about which parameters to hold fixed, and which to allow to vary in the theory. These choices affect the functional forms of the thermodynamic functions, which will affect $C_V$. Equation (\ref{CVvz}) maintains its form despite these different choices.

\section{Proof of the Alternate Condition for Criticality}
\label{appendixcritical}
We provide here the proof of the condition presented in (\ref{criticalpt2}). We noted when discussing criticality that the equations we wish to solve cannot be acquired analytically. Nonetheless, we noticed that at certain values of $\nu$, $C_p (T)$ begins developing a singularity at some value of $T$. This is strong evidence of a first order phase transition, since singularities in the heat capacities directly imply a jump discontinuity in the entropy, assuming the temperature function is finite there. This jump discontinuity in the entropy is the tell-tale sign of a first order phase transition.

We thus seek the critical point by seeking divergences in the entropy as a function of temperature. At fixed $\nu$, we may regard this curve as a vector-valued function, parameterized by $z$, given by
\begin{equation}
    \label{entropyvector}
    \Vec{s}(z)=\biggl\langle T(\nu,z), S(\nu,z) \biggr\rangle.
\end{equation}
The vector tangent to this curve for each value of $z$, assuming the relevant derivatives exist, is given by
\begin{equation}
    \label{entropyderivvector}
    \frac{d \Vec{s}}{dz}=\biggl\langle \frac{dT}{dz}, \frac{dS}{dz} \biggr\rangle.
\end{equation}

Based on the geometry of this function for various values of $\nu$, we surmised that at the jump discontinuity we should expect this vector to be pointing straight down. That is, this vector must be proportional to the vector $-\hat{y}=\langle 0,-1 \rangle$ at the jump discontinuity. That gives us two conditions the jump discontinuity must obey:
\begin{equation}
    \label{conditions}
    \frac{dT}{dz}=0,\; \text{and}\;\;\;\frac{dS}{dz}< 0.
\end{equation}
We must, however, be slightly more stringent. To ensure that this discontinuity will produce the type of divergence expected in the heat capacity, the curve connecting the two disparate branches must have no concavity. It must be a vertical line. To satisfy this requirement, we must also demand that the curvature of this curve at the point satisfying (\ref{conditions}) be zero, since that is the curvature of a straight line. Recall that for any regular, simply connected curve parameterized by arclength, one may define the Frenet-Serret frame, a set of unit vectors corresponding to the tangent, normal, and binormal directions at each point of the curve, given by $\Vec{t}, \Vec{n},$ and~$\Vec{b}$, respectively. Since these vectors form a linearly independent set of vectors, one may differentiate these vectors and express the results in terms of the original vectors themselves. The result relevant to our discussion is
\begin{equation}
    \label{curvature}
    \frac{d\Vec{t}}{ds}=K(s)\Vec{n},
\end{equation}
 where $K(s)$ is the curvature of the curve at any point. Thus, we seek that the derivative of the tangent to the curve \emph{when parameterized by arclength} be zero. We therefore normalize (\ref{entropyderivvector}) and then differentiate with respect to $z$. Letting primes stand for $z$ derivatives, the result is
\begin{widetext}
\begin{equation}
    \label{2ndcondition}
    \frac{d\Vec{t}}{dz}=\biggl\langle\frac{(T'^2+S'^2) T''-T' (T' T''+S' S'')}{(T'^2+S'^2)^{3/2}}, \frac{(T'^2+S'^2) S''-S' (T' T''+S' S'')}{(T'^2+S'^2)^{3/2}} \biggr\rangle.
\end{equation}
\end{widetext}
Each component of (\ref{2ndcondition}) must be zero. Plugging in the conditions $T'=0$ and $T''=0$ indeed makes them both zero, while $S'\neq 0$ prevents any undefined or indeterminate points from arising.

\section{The Exact Value of $T_{min}$ for $V=V_c$}
\label{appendixTmin}
The numerical plots for $C_V(T)$ display two interesting characteristics for every fixed volume. First, the plots contain two branches, both of which diverge at some finite temperature, $T_{div}$. Furthermore, one of these branches never reaches $T=0$, but rather begins at some $T_{min}>0$. We can calculate the latter exactly for the critical volume.

To calculate $T_{\rm min}$, we calculate $\lim_{z\to\infty} T(\nu(z),z)$, where $\nu=\nu(z)$ is the solution of $V(\nu,z)=0$ going through the critical point $(\nu_c, z_c)$. One way to write the result is
\begin{widetext}
\begin{equation}
        \label{Tmin}
T_{\rm min}=\frac{68\sqrt{10}\cos{(\gamma)}+62\cos{(2\gamma)}+5\cos{(4\gamma)}-6\sqrt{1110}\sin{(\gamma)}+36\sqrt{111}\sin{(2\gamma)}+15\sqrt{111}\sin{(4\gamma)}-90}{3\pi (104\sqrt{10}\cos{(\gamma)}+10\cos({2\gamma)}+12\sqrt{1110}\sin{(\gamma)}+30\sqrt{111}\sin{(2\gamma)}+525)}\ ,    
    \end{equation}
    \end{widetext}
where $\gamma\equiv \frac{1}{3}\arctan{(3 \sqrt{111})}$.
Evaluating this expression numerically yields $T_{\rm min}\simeq{0.0408}$, as stated in the body of this paper.

\bibliographystyle{apsrev4-1}
\bibliography{citations}

\begin{thebibliography}{40}%
\makeatletter
\providecommand \@ifxundefined [1]{%
 \@ifx{#1\undefined}
}%
\providecommand \@ifnum [1]{%
 \ifnum #1\expandafter \@firstoftwo
 \else \expandafter \@secondoftwo
 \fi
}%
\providecommand \@ifx [1]{%
 \ifx #1\expandafter \@firstoftwo
 \else \expandafter \@secondoftwo
 \fi
}%
\providecommand \natexlab [1]{#1}%
\providecommand \enquote  [1]{``#1''}%
\providecommand \bibnamefont  [1]{#1}%
\providecommand \bibfnamefont [1]{#1}%
\providecommand \citenamefont [1]{#1}%
\providecommand \href@noop [0]{\@secondoftwo}%
\providecommand \href [0]{\begingroup \@sanitize@url \@href}%
\providecommand \@href[1]{\@@startlink{#1}\@@href}%
\providecommand \@@href[1]{\endgroup#1\@@endlink}%
\providecommand \@sanitize@url [0]{\catcode `\\12\catcode `\$12\catcode `\&12\catcode `\#12\catcode `\^12\catcode `\_12\catcode `\%12\relax}%
\providecommand \@@startlink[1]{}%
\providecommand \@@endlink[0]{}%
\providecommand \url  [0]{\begingroup\@sanitize@url \@url }%
\providecommand \@url [1]{\endgroup\@href {#1}{\urlprefix }}%
\providecommand \urlprefix  [0]{URL }%
\providecommand \Eprint [0]{\href }%
\providecommand \doibase [0]{http://dx.doi.org/}%
\providecommand \selectlanguage [0]{\@gobble}%
\providecommand \bibinfo  [0]{\@secondoftwo}%
\providecommand \bibfield  [0]{\@secondoftwo}%
\providecommand \translation [1]{[#1]}%
\providecommand \BibitemOpen [0]{}%
\providecommand \bibitemStop [0]{}%
\providecommand \bibitemNoStop [0]{.\EOS\space}%
\providecommand \EOS [0]{\spacefactor3000\relax}%
\providecommand \BibitemShut  [1]{\csname bibitem#1\endcsname}%
\let\auto@bib@innerbib\@empty
\bibitem [{\citenamefont {Bekenstein}(1973)}]{Bekenstein:1973ur}%
  \BibitemOpen
  \bibfield  {author} {\bibinfo {author} {\bibfnamefont {J.~D.}\ \bibnamefont {Bekenstein}},\ }\href {\doibase 10.1103/PhysRevD.7.2333} {\bibfield  {journal} {\bibinfo  {journal} {Phys. Rev. D}\ }\textbf {\bibinfo {volume} {7}},\ \bibinfo {pages} {2333} (\bibinfo {year} {1973})}\BibitemShut {NoStop}%
\bibitem [{\citenamefont {Bekenstein}(1974)}]{Bekenstein:1974ax}%
  \BibitemOpen
  \bibfield  {author} {\bibinfo {author} {\bibfnamefont {J.~D.}\ \bibnamefont {Bekenstein}},\ }\href {\doibase 10.1103/PhysRevD.9.3292} {\bibfield  {journal} {\bibinfo  {journal} {Phys. Rev. D}\ }\textbf {\bibinfo {volume} {9}},\ \bibinfo {pages} {3292} (\bibinfo {year} {1974})}\BibitemShut {NoStop}%
\bibitem [{\citenamefont {Hawking}(1975)}]{Hawking:1975vcx}%
  \BibitemOpen
  \bibfield  {author} {\bibinfo {author} {\bibfnamefont {S.~W.}\ \bibnamefont {Hawking}},\ }\href {\doibase 10.1007/BF02345020} {\bibfield  {journal} {\bibinfo  {journal} {Commun. Math. Phys.}\ }\textbf {\bibinfo {volume} {43}},\ \bibinfo {pages} {199} (\bibinfo {year} {1975})},\ \bibinfo {note} {[Erratum: Commun.Math.Phys. 46, 206 (1976)]}\BibitemShut {NoStop}%
\bibitem [{\citenamefont {Hawking}(1976)}]{Hawking:1976de}%
  \BibitemOpen
  \bibfield  {author} {\bibinfo {author} {\bibfnamefont {S.~W.}\ \bibnamefont {Hawking}},\ }\href {\doibase 10.1103/PhysRevD.13.191} {\bibfield  {journal} {\bibinfo  {journal} {Phys. Rev. D}\ }\textbf {\bibinfo {volume} {13}},\ \bibinfo {pages} {191} (\bibinfo {year} {1976})}\BibitemShut {NoStop}%
\bibitem [{\citenamefont {Bardeen}\ \emph {et~al.}(1973)\citenamefont {Bardeen}, \citenamefont {Carter},\ and\ \citenamefont {Hawking}}]{Bardeen:1973gs}%
  \BibitemOpen
  \bibfield  {author} {\bibinfo {author} {\bibfnamefont {J.~M.}\ \bibnamefont {Bardeen}}, \bibinfo {author} {\bibfnamefont {B.}~\bibnamefont {Carter}}, \ and\ \bibinfo {author} {\bibfnamefont {S.~W.}\ \bibnamefont {Hawking}},\ }\href {\doibase 10.1007/BF01645742} {\bibfield  {journal} {\bibinfo  {journal} {Commun. Math. Phys.}\ }\textbf {\bibinfo {volume} {31}},\ \bibinfo {pages} {161} (\bibinfo {year} {1973})}\BibitemShut {NoStop}%
\bibitem [{\citenamefont {Hawking}\ and\ \citenamefont {Page}(1983)}]{Hawking:1982dh}%
  \BibitemOpen
  \bibfield  {author} {\bibinfo {author} {\bibfnamefont {S.~W.}\ \bibnamefont {Hawking}}\ and\ \bibinfo {author} {\bibfnamefont {D.~N.}\ \bibnamefont {Page}},\ }\href {\doibase 10.1007/BF01208266} {\bibfield  {journal} {\bibinfo  {journal} {Commun. Math. Phys.}\ }\textbf {\bibinfo {volume} {87}},\ \bibinfo {pages} {577} (\bibinfo {year} {1983})}\BibitemShut {NoStop}%
\bibitem [{\citenamefont {Henneaux}\ and\ \citenamefont {Teitelboim}(1984)}]{Henneaux:1984ji}%
  \BibitemOpen
  \bibfield  {author} {\bibinfo {author} {\bibfnamefont {M.}~\bibnamefont {Henneaux}}\ and\ \bibinfo {author} {\bibfnamefont {C.}~\bibnamefont {Teitelboim}},\ }\href {\doibase 10.1016/0370-2693(84)91493-X} {\bibfield  {journal} {\bibinfo  {journal} {Phys. Lett. B}\ }\textbf {\bibinfo {volume} {143}},\ \bibinfo {pages} {415} (\bibinfo {year} {1984})}\BibitemShut {NoStop}%
\bibitem [{\citenamefont {Teitelboim}(1985)}]{Teitelboim:1985dp}%
  \BibitemOpen
  \bibfield  {author} {\bibinfo {author} {\bibfnamefont {C.}~\bibnamefont {Teitelboim}},\ }\href {\doibase 10.1016/0370-2693(85)91186-4} {\bibfield  {journal} {\bibinfo  {journal} {Phys. Lett. B}\ }\textbf {\bibinfo {volume} {158}},\ \bibinfo {pages} {293} (\bibinfo {year} {1985})}\BibitemShut {NoStop}%
\bibitem [{\citenamefont {Henneaux}\ and\ \citenamefont {Teitelboim}(1989)}]{Henneaux:1989zc}%
  \BibitemOpen
  \bibfield  {author} {\bibinfo {author} {\bibfnamefont {M.}~\bibnamefont {Henneaux}}\ and\ \bibinfo {author} {\bibfnamefont {C.}~\bibnamefont {Teitelboim}},\ }\href {\doibase 10.1016/0370-2693(89)91251-3} {\bibfield  {journal} {\bibinfo  {journal} {Phys. Lett. B}\ }\textbf {\bibinfo {volume} {222}},\ \bibinfo {pages} {195} (\bibinfo {year} {1989})}\BibitemShut {NoStop}%
\bibitem [{\citenamefont {Kastor}\ \emph {et~al.}(2009)\citenamefont {Kastor}, \citenamefont {Ray},\ and\ \citenamefont {Traschen}}]{Kastor:2009wy}%
  \BibitemOpen
  \bibfield  {author} {\bibinfo {author} {\bibfnamefont {D.}~\bibnamefont {Kastor}}, \bibinfo {author} {\bibfnamefont {S.}~\bibnamefont {Ray}}, \ and\ \bibinfo {author} {\bibfnamefont {J.}~\bibnamefont {Traschen}},\ }\href {\doibase 10.1088/0264-9381/26/19/195011} {\bibfield  {journal} {\bibinfo  {journal} {Class. Quant. Grav.}\ }\textbf {\bibinfo {volume} {26}},\ \bibinfo {pages} {195011} (\bibinfo {year} {2009})},\ \Eprint {http://arxiv.org/abs/0904.2765} {arXiv:0904.2765 [hep-th]} \BibitemShut {NoStop}%
\bibitem [{\citenamefont {Cvetic}\ \emph {et~al.}(2011)\citenamefont {Cvetic}, \citenamefont {Gibbons}, \citenamefont {Kubiznak},\ and\ \citenamefont {Pope}}]{Cvetic:2010jb}%
  \BibitemOpen
  \bibfield  {author} {\bibinfo {author} {\bibfnamefont {M.}~\bibnamefont {Cvetic}}, \bibinfo {author} {\bibfnamefont {G.~W.}\ \bibnamefont {Gibbons}}, \bibinfo {author} {\bibfnamefont {D.}~\bibnamefont {Kubiznak}}, \ and\ \bibinfo {author} {\bibfnamefont {C.~N.}\ \bibnamefont {Pope}},\ }\href {\doibase 10.1103/PhysRevD.84.024037} {\bibfield  {journal} {\bibinfo  {journal} {Phys. Rev. D}\ }\textbf {\bibinfo {volume} {84}},\ \bibinfo {pages} {024037} (\bibinfo {year} {2011})},\ \Eprint {http://arxiv.org/abs/1012.2888} {arXiv:1012.2888 [hep-th]} \BibitemShut {NoStop}%
\bibitem [{\citenamefont {Witten}(1998)}]{Witten:1998zw}%
  \BibitemOpen
  \bibfield  {author} {\bibinfo {author} {\bibfnamefont {E.}~\bibnamefont {Witten}},\ }\href {\doibase 10.4310/ATMP.1998.v2.n3.a3} {\bibfield  {journal} {\bibinfo  {journal} {Adv. Theor. Math. Phys.}\ }\textbf {\bibinfo {volume} {2}},\ \bibinfo {pages} {505} (\bibinfo {year} {1998})},\ \Eprint {http://arxiv.org/abs/hep-th/9803131} {arXiv:hep-th/9803131} \BibitemShut {NoStop}%
\bibitem [{\citenamefont {Dolan}(2011)}]{Dolan:2010ha}%
  \BibitemOpen
  \bibfield  {author} {\bibinfo {author} {\bibfnamefont {B.~P.}\ \bibnamefont {Dolan}},\ }\href {\doibase 10.1088/0264-9381/28/12/125020} {\bibfield  {journal} {\bibinfo  {journal} {Class. Quant. Grav.}\ }\textbf {\bibinfo {volume} {28}},\ \bibinfo {pages} {125020} (\bibinfo {year} {2011})},\ \Eprint {http://arxiv.org/abs/1008.5023} {arXiv:1008.5023 [gr-qc]} \BibitemShut {NoStop}%
\bibitem [{\citenamefont {Kubiznak}\ \emph {et~al.}(2017)\citenamefont {Kubiznak}, \citenamefont {Mann},\ and\ \citenamefont {Teo}}]{Kubiznak:2016qmn}%
  \BibitemOpen
  \bibfield  {author} {\bibinfo {author} {\bibfnamefont {D.}~\bibnamefont {Kubiznak}}, \bibinfo {author} {\bibfnamefont {R.~B.}\ \bibnamefont {Mann}}, \ and\ \bibinfo {author} {\bibfnamefont {M.}~\bibnamefont {Teo}},\ }\href {\doibase 10.1088/1361-6382/aa5c69} {\bibfield  {journal} {\bibinfo  {journal} {Class. Quant. Grav.}\ }\textbf {\bibinfo {volume} {34}},\ \bibinfo {pages} {063001} (\bibinfo {year} {2017})},\ \Eprint {http://arxiv.org/abs/1608.06147} {arXiv:1608.06147 [hep-th]} \BibitemShut {NoStop}%
\bibitem [{\citenamefont {Johnson}(2020{\natexlab{a}})}]{Johnson:2019vqf}%
  \BibitemOpen
  \bibfield  {author} {\bibinfo {author} {\bibfnamefont {C.~V.}\ \bibnamefont {Johnson}},\ }\href {\doibase 10.1088/1361-6382/ab685a} {\bibfield  {journal} {\bibinfo  {journal} {Class. Quant. Grav.}\ }\textbf {\bibinfo {volume} {37}},\ \bibinfo {pages} {054003} (\bibinfo {year} {2020}{\natexlab{a}})},\ \Eprint {http://arxiv.org/abs/1905.00539} {arXiv:1905.00539 [hep-th]} \BibitemShut {NoStop}%
\bibitem [{\citenamefont {Chamblin}\ \emph {et~al.}(1999)\citenamefont {Chamblin}, \citenamefont {Emparan}, \citenamefont {Johnson},\ and\ \citenamefont {Myers}}]{Chamblin:1998pz}%
  \BibitemOpen
  \bibfield  {author} {\bibinfo {author} {\bibfnamefont {A.}~\bibnamefont {Chamblin}}, \bibinfo {author} {\bibfnamefont {R.}~\bibnamefont {Emparan}}, \bibinfo {author} {\bibfnamefont {C.~V.}\ \bibnamefont {Johnson}}, \ and\ \bibinfo {author} {\bibfnamefont {R.~C.}\ \bibnamefont {Myers}},\ }\href {\doibase 10.1103/PhysRevD.59.064010} {\bibfield  {journal} {\bibinfo  {journal} {Phys. Rev. D}\ }\textbf {\bibinfo {volume} {59}},\ \bibinfo {pages} {064010} (\bibinfo {year} {1999})},\ \Eprint {http://arxiv.org/abs/hep-th/9808177} {arXiv:hep-th/9808177} \BibitemShut {NoStop}%
\bibitem [{\citenamefont {Kubiznak}\ and\ \citenamefont {Mann}(2015)}]{Kubiznak:2014zwa}%
  \BibitemOpen
  \bibfield  {author} {\bibinfo {author} {\bibfnamefont {D.}~\bibnamefont {Kubiznak}}\ and\ \bibinfo {author} {\bibfnamefont {R.~B.}\ \bibnamefont {Mann}},\ }\href {\doibase 10.1139/cjp-2014-0465} {\bibfield  {journal} {\bibinfo  {journal} {Can. J. Phys.}\ }\textbf {\bibinfo {volume} {93}},\ \bibinfo {pages} {999} (\bibinfo {year} {2015})},\ \Eprint {http://arxiv.org/abs/1404.2126} {arXiv:1404.2126 [gr-qc]} \BibitemShut {NoStop}%
\bibitem [{\citenamefont {Banados}\ \emph {et~al.}(1992)\citenamefont {Banados}, \citenamefont {Teitelboim},\ and\ \citenamefont {Zanelli}}]{Banados:1992wn}%
  \BibitemOpen
  \bibfield  {author} {\bibinfo {author} {\bibfnamefont {M.}~\bibnamefont {Banados}}, \bibinfo {author} {\bibfnamefont {C.}~\bibnamefont {Teitelboim}}, \ and\ \bibinfo {author} {\bibfnamefont {J.}~\bibnamefont {Zanelli}},\ }\href {\doibase 10.1103/PhysRevLett.69.1849} {\bibfield  {journal} {\bibinfo  {journal} {Phys. Rev. Lett.}\ }\textbf {\bibinfo {volume} {69}},\ \bibinfo {pages} {1849} (\bibinfo {year} {1992})},\ \Eprint {http://arxiv.org/abs/hep-th/9204099} {arXiv:hep-th/9204099} \BibitemShut {NoStop}%
\bibitem [{\citenamefont {Banados}\ \emph {et~al.}(1993)\citenamefont {Banados}, \citenamefont {Henneaux}, \citenamefont {Teitelboim},\ and\ \citenamefont {Zanelli}}]{Banados:1992gq}%
  \BibitemOpen
  \bibfield  {author} {\bibinfo {author} {\bibfnamefont {M.}~\bibnamefont {Banados}}, \bibinfo {author} {\bibfnamefont {M.}~\bibnamefont {Henneaux}}, \bibinfo {author} {\bibfnamefont {C.}~\bibnamefont {Teitelboim}}, \ and\ \bibinfo {author} {\bibfnamefont {J.}~\bibnamefont {Zanelli}},\ }\href {\doibase 10.1103/PhysRevD.48.1506} {\bibfield  {journal} {\bibinfo  {journal} {Phys. Rev. D}\ }\textbf {\bibinfo {volume} {48}},\ \bibinfo {pages} {1506} (\bibinfo {year} {1993})},\ \bibinfo {note} {[Erratum: Phys.Rev.D 88, 069902 (2013)]},\ \Eprint {http://arxiv.org/abs/gr-qc/9302012} {arXiv:gr-qc/9302012} \BibitemShut {NoStop}%
\bibitem [{\citenamefont {Frassino}\ \emph {et~al.}(2015)\citenamefont {Frassino}, \citenamefont {Mann},\ and\ \citenamefont {Mureika}}]{Frassino:2015oca}%
  \BibitemOpen
  \bibfield  {author} {\bibinfo {author} {\bibfnamefont {A.~M.}\ \bibnamefont {Frassino}}, \bibinfo {author} {\bibfnamefont {R.~B.}\ \bibnamefont {Mann}}, \ and\ \bibinfo {author} {\bibfnamefont {J.~R.}\ \bibnamefont {Mureika}},\ }\href {\doibase 10.1103/PhysRevD.92.124069} {\bibfield  {journal} {\bibinfo  {journal} {Phys. Rev. D}\ }\textbf {\bibinfo {volume} {92}},\ \bibinfo {pages} {124069} (\bibinfo {year} {2015})},\ \Eprint {http://arxiv.org/abs/1509.05481} {arXiv:1509.05481 [gr-qc]} \BibitemShut {NoStop}%
\bibitem [{\citenamefont {Johnson}(2020{\natexlab{b}})}]{Johnson:2019mdp}%
  \BibitemOpen
  \bibfield  {author} {\bibinfo {author} {\bibfnamefont {C.~V.}\ \bibnamefont {Johnson}},\ }\href {\doibase 10.1142/S0217732320500984} {\bibfield  {journal} {\bibinfo  {journal} {Mod. Phys. Lett. A}\ }\textbf {\bibinfo {volume} {35}},\ \bibinfo {pages} {2050098} (\bibinfo {year} {2020}{\natexlab{b}})},\ \Eprint {http://arxiv.org/abs/1906.00993} {arXiv:1906.00993 [hep-th]} \BibitemShut {NoStop}%
\bibitem [{\citenamefont {Johnson}\ \emph {et~al.}(2020)\citenamefont {Johnson}, \citenamefont {Martin},\ and\ \citenamefont {Svesko}}]{Johnson:2019wcq}%
  \BibitemOpen
  \bibfield  {author} {\bibinfo {author} {\bibfnamefont {C.~V.}\ \bibnamefont {Johnson}}, \bibinfo {author} {\bibfnamefont {V.~L.}\ \bibnamefont {Martin}}, \ and\ \bibinfo {author} {\bibfnamefont {A.}~\bibnamefont {Svesko}},\ }\href {\doibase 10.1103/PhysRevD.101.086006} {\bibfield  {journal} {\bibinfo  {journal} {Phys. Rev. D}\ }\textbf {\bibinfo {volume} {101}},\ \bibinfo {pages} {086006} (\bibinfo {year} {2020})},\ \Eprint {http://arxiv.org/abs/1911.05286} {arXiv:1911.05286 [hep-th]} \BibitemShut {NoStop}%
\bibitem [{\citenamefont {Cong}\ and\ \citenamefont {Mann}(2019)}]{Cong:2019bud}%
  \BibitemOpen
  \bibfield  {author} {\bibinfo {author} {\bibfnamefont {W.}~\bibnamefont {Cong}}\ and\ \bibinfo {author} {\bibfnamefont {R.~B.}\ \bibnamefont {Mann}},\ }\href {\doibase 10.1007/JHEP11(2019)004} {\bibfield  {journal} {\bibinfo  {journal} {JHEP}\ }\textbf {\bibinfo {volume} {11}},\ \bibinfo {pages} {004} (\bibinfo {year} {2019})},\ \Eprint {http://arxiv.org/abs/1908.01254} {arXiv:1908.01254 [gr-qc]} \BibitemShut {NoStop}%
\bibitem [{\citenamefont {Song}\ \emph {et~al.}(2023)\citenamefont {Song}, \citenamefont {He},\ and\ \citenamefont {Mu}}]{Song:2023zre}%
  \BibitemOpen
  \bibfield  {author} {\bibinfo {author} {\bibfnamefont {Y.}~\bibnamefont {Song}}, \bibinfo {author} {\bibfnamefont {Y.}~\bibnamefont {He}}, \ and\ \bibinfo {author} {\bibfnamefont {B.}~\bibnamefont {Mu}},\ }\href@noop {} {\  (\bibinfo {year} {2023})},\ \Eprint {http://arxiv.org/abs/2306.01030} {arXiv:2306.01030 [gr-qc]} \BibitemShut {NoStop}%
\bibitem [{\citenamefont {Jing}\ \emph {et~al.}(2021)\citenamefont {Jing}, \citenamefont {Mu}, \citenamefont {Tao},\ and\ \citenamefont {Wang}}]{Jing:2020sdf}%
  \BibitemOpen
  \bibfield  {author} {\bibinfo {author} {\bibfnamefont {H.}~\bibnamefont {Jing}}, \bibinfo {author} {\bibfnamefont {B.}~\bibnamefont {Mu}}, \bibinfo {author} {\bibfnamefont {J.}~\bibnamefont {Tao}}, \ and\ \bibinfo {author} {\bibfnamefont {P.}~\bibnamefont {Wang}},\ }\href {\doibase 10.1088/1674-1137/abf1dc} {\bibfield  {journal} {\bibinfo  {journal} {Chin. Phys. C}\ }\textbf {\bibinfo {volume} {45}},\ \bibinfo {pages} {065103} (\bibinfo {year} {2021})},\ \Eprint {http://arxiv.org/abs/2012.14206} {arXiv:2012.14206 [gr-qc]} \BibitemShut {NoStop}%
\bibitem [{\citenamefont {Jahani~Poshteh}\ and\ \citenamefont {Mann}(2021)}]{JahaniPoshteh:2021clv}%
  \BibitemOpen
  \bibfield  {author} {\bibinfo {author} {\bibfnamefont {M.~B.}\ \bibnamefont {Jahani~Poshteh}}\ and\ \bibinfo {author} {\bibfnamefont {R.~B.}\ \bibnamefont {Mann}},\ }\href {\doibase 10.1103/PhysRevD.103.104024} {\bibfield  {journal} {\bibinfo  {journal} {Phys. Rev. D}\ }\textbf {\bibinfo {volume} {103}},\ \bibinfo {pages} {104024} (\bibinfo {year} {2021})},\ \Eprint {http://arxiv.org/abs/2103.04365} {arXiv:2103.04365 [hep-th]} \BibitemShut {NoStop}%
\bibitem [{\citenamefont {Song}\ and\ \citenamefont {Mu}(2023)}]{Song:2023kvq}%
  \BibitemOpen
  \bibfield  {author} {\bibinfo {author} {\bibfnamefont {Y.}~\bibnamefont {Song}}\ and\ \bibinfo {author} {\bibfnamefont {B.}~\bibnamefont {Mu}},\ }\href@noop {} {\  (\bibinfo {year} {2023})},\ \Eprint {http://arxiv.org/abs/2304.09760} {arXiv:2304.09760 [gr-qc]} \BibitemShut {NoStop}%
\bibitem [{\citenamefont {He}\ and\ \citenamefont {Mu}(2023)}]{He:2023tiz}%
  \BibitemOpen
  \bibfield  {author} {\bibinfo {author} {\bibfnamefont {Y.}~\bibnamefont {He}}\ and\ \bibinfo {author} {\bibfnamefont {B.}~\bibnamefont {Mu}},\ }\href@noop {} {\  (\bibinfo {year} {2023})},\ \Eprint {http://arxiv.org/abs/2305.02196} {arXiv:2305.02196 [gr-qc]} \BibitemShut {NoStop}%
\bibitem [{\citenamefont {Appels}\ \emph {et~al.}(2020)\citenamefont {Appels}, \citenamefont {Cuspinera}, \citenamefont {Gregory}, \citenamefont {Krtou\v{s}},\ and\ \citenamefont {Kubiz\v{n}\'ak}}]{Appels:2019vow}%
  \BibitemOpen
  \bibfield  {author} {\bibinfo {author} {\bibfnamefont {M.}~\bibnamefont {Appels}}, \bibinfo {author} {\bibfnamefont {L.}~\bibnamefont {Cuspinera}}, \bibinfo {author} {\bibfnamefont {R.}~\bibnamefont {Gregory}}, \bibinfo {author} {\bibfnamefont {P.}~\bibnamefont {Krtou\v{s}}}, \ and\ \bibinfo {author} {\bibfnamefont {D.}~\bibnamefont {Kubiz\v{n}\'ak}},\ }\href {\doibase 10.1007/JHEP02(2020)195} {\bibfield  {journal} {\bibinfo  {journal} {JHEP}\ }\textbf {\bibinfo {volume} {02}},\ \bibinfo {pages} {195} (\bibinfo {year} {2020})},\ \Eprint {http://arxiv.org/abs/1911.12817} {arXiv:1911.12817 [hep-th]} \BibitemShut {NoStop}%
\bibitem [{\citenamefont {Emparan}\ \emph {et~al.}(2000)\citenamefont {Emparan}, \citenamefont {Horowitz},\ and\ \citenamefont {Myers}}]{Emparan:1999fd}%
  \BibitemOpen
  \bibfield  {author} {\bibinfo {author} {\bibfnamefont {R.}~\bibnamefont {Emparan}}, \bibinfo {author} {\bibfnamefont {G.~T.}\ \bibnamefont {Horowitz}}, \ and\ \bibinfo {author} {\bibfnamefont {R.~C.}\ \bibnamefont {Myers}},\ }\href {\doibase 10.1088/1126-6708/2000/01/021} {\bibfield  {journal} {\bibinfo  {journal} {JHEP}\ }\textbf {\bibinfo {volume} {01}},\ \bibinfo {pages} {021} (\bibinfo {year} {2000})},\ \Eprint {http://arxiv.org/abs/hep-th/9912135} {arXiv:hep-th/9912135} \BibitemShut {NoStop}%
\bibitem [{\citenamefont {Emparan}\ \emph {et~al.}(2020)\citenamefont {Emparan}, \citenamefont {Frassino},\ and\ \citenamefont {Way}}]{Emparan:2020znc}%
  \BibitemOpen
  \bibfield  {author} {\bibinfo {author} {\bibfnamefont {R.}~\bibnamefont {Emparan}}, \bibinfo {author} {\bibfnamefont {A.~M.}\ \bibnamefont {Frassino}}, \ and\ \bibinfo {author} {\bibfnamefont {B.}~\bibnamefont {Way}},\ }\href {\doibase 10.1007/JHEP11(2020)137} {\bibfield  {journal} {\bibinfo  {journal} {JHEP}\ }\textbf {\bibinfo {volume} {11}},\ \bibinfo {pages} {137} (\bibinfo {year} {2020})},\ \Eprint {http://arxiv.org/abs/2007.15999} {arXiv:2007.15999 [hep-th]} \BibitemShut {NoStop}%
\bibitem [{\citenamefont {Frassino}\ \emph {et~al.}(2023)\citenamefont {Frassino}, \citenamefont {Pedraza}, \citenamefont {Svesko},\ and\ \citenamefont {Visser}}]{Frassino:2022zaz}%
  \BibitemOpen
  \bibfield  {author} {\bibinfo {author} {\bibfnamefont {A.~M.}\ \bibnamefont {Frassino}}, \bibinfo {author} {\bibfnamefont {J.~F.}\ \bibnamefont {Pedraza}}, \bibinfo {author} {\bibfnamefont {A.}~\bibnamefont {Svesko}}, \ and\ \bibinfo {author} {\bibfnamefont {M.~R.}\ \bibnamefont {Visser}},\ }\href {\doibase 10.1103/PhysRevLett.130.161501} {\bibfield  {journal} {\bibinfo  {journal} {Phys. Rev. Lett.}\ }\textbf {\bibinfo {volume} {130}},\ \bibinfo {pages} {161501} (\bibinfo {year} {2023})},\ \Eprint {http://arxiv.org/abs/2212.14055} {arXiv:2212.14055 [hep-th]} \BibitemShut {NoStop}%
\bibitem [{\citenamefont {Karch}\ and\ \citenamefont {Randall}(2001)}]{Karch:2000ct}%
  \BibitemOpen
  \bibfield  {author} {\bibinfo {author} {\bibfnamefont {A.}~\bibnamefont {Karch}}\ and\ \bibinfo {author} {\bibfnamefont {L.}~\bibnamefont {Randall}},\ }\href {\doibase 10.1088/1126-6708/2001/05/008} {\bibfield  {journal} {\bibinfo  {journal} {JHEP}\ }\textbf {\bibinfo {volume} {05}},\ \bibinfo {pages} {008} (\bibinfo {year} {2001})},\ \Eprint {http://arxiv.org/abs/hep-th/0011156} {arXiv:hep-th/0011156} \BibitemShut {NoStop}%
\bibitem [{\citenamefont {Frassino}\ \emph {et~al.}(2024)\citenamefont {Frassino}, \citenamefont {Pedraza}, \citenamefont {Svesko},\ and\ \citenamefont {Visser}}]{Frassino:2023wpc}%
  \BibitemOpen
  \bibfield  {author} {\bibinfo {author} {\bibfnamefont {A.~M.}\ \bibnamefont {Frassino}}, \bibinfo {author} {\bibfnamefont {J.~F.}\ \bibnamefont {Pedraza}}, \bibinfo {author} {\bibfnamefont {A.}~\bibnamefont {Svesko}}, \ and\ \bibinfo {author} {\bibfnamefont {M.~R.}\ \bibnamefont {Visser}},\ }\href {\doibase 10.1103/PhysRevD.109.124040} {\bibfield  {journal} {\bibinfo  {journal} {Phys. Rev. D}\ }\textbf {\bibinfo {volume} {109}},\ \bibinfo {pages} {124040} (\bibinfo {year} {2024})},\ \Eprint {http://arxiv.org/abs/2310.12220} {arXiv:2310.12220 [hep-th]} \BibitemShut {NoStop}%
\bibitem [{\citenamefont {Emparan}\ \emph {et~al.}(2002)\citenamefont {Emparan}, \citenamefont {Fabbri},\ and\ \citenamefont {Kaloper}}]{Emparan:2002px}%
  \BibitemOpen
  \bibfield  {author} {\bibinfo {author} {\bibfnamefont {R.}~\bibnamefont {Emparan}}, \bibinfo {author} {\bibfnamefont {A.}~\bibnamefont {Fabbri}}, \ and\ \bibinfo {author} {\bibfnamefont {N.}~\bibnamefont {Kaloper}},\ }\href {\doibase 10.1088/1126-6708/2002/08/043} {\bibfield  {journal} {\bibinfo  {journal} {JHEP}\ }\textbf {\bibinfo {volume} {08}},\ \bibinfo {pages} {043} (\bibinfo {year} {2002})},\ \Eprint {http://arxiv.org/abs/hep-th/0206155} {arXiv:hep-th/0206155} \BibitemShut {NoStop}%
\bibitem [{\citenamefont {Kudoh}\ and\ \citenamefont {Kurita}(2004)}]{Kudoh:2004ub}%
  \BibitemOpen
  \bibfield  {author} {\bibinfo {author} {\bibfnamefont {H.}~\bibnamefont {Kudoh}}\ and\ \bibinfo {author} {\bibfnamefont {Y.}~\bibnamefont {Kurita}},\ }\href {\doibase 10.1103/PhysRevD.70.084029} {\bibfield  {journal} {\bibinfo  {journal} {Phys. Rev. D}\ }\textbf {\bibinfo {volume} {70}},\ \bibinfo {pages} {084029} (\bibinfo {year} {2004})},\ \Eprint {http://arxiv.org/abs/gr-qc/0406107} {arXiv:gr-qc/0406107} \BibitemShut {NoStop}%
\bibitem [{\citenamefont {Kubiznak}\ and\ \citenamefont {Mann}(2012)}]{Kubiznak:2012wp}%
  \BibitemOpen
  \bibfield  {author} {\bibinfo {author} {\bibfnamefont {D.}~\bibnamefont {Kubiznak}}\ and\ \bibinfo {author} {\bibfnamefont {R.~B.}\ \bibnamefont {Mann}},\ }\href {\doibase 10.1007/JHEP07(2012)033} {\bibfield  {journal} {\bibinfo  {journal} {JHEP}\ }\textbf {\bibinfo {volume} {07}},\ \bibinfo {pages} {033} (\bibinfo {year} {2012})},\ \Eprint {http://arxiv.org/abs/1205.0559} {arXiv:1205.0559 [hep-th]} \BibitemShut {NoStop}%
\bibitem [{\citenamefont {Altamirano}\ \emph {et~al.}(2013)\citenamefont {Altamirano}, \citenamefont {Kubiznak},\ and\ \citenamefont {Mann}}]{Altamirano:2013ane}%
  \BibitemOpen
  \bibfield  {author} {\bibinfo {author} {\bibfnamefont {N.}~\bibnamefont {Altamirano}}, \bibinfo {author} {\bibfnamefont {D.}~\bibnamefont {Kubiznak}}, \ and\ \bibinfo {author} {\bibfnamefont {R.~B.}\ \bibnamefont {Mann}},\ }\href {\doibase 10.1103/PhysRevD.88.101502} {\bibfield  {journal} {\bibinfo  {journal} {Phys. Rev. D}\ }\textbf {\bibinfo {volume} {88}},\ \bibinfo {pages} {101502} (\bibinfo {year} {2013})},\ \Eprint {http://arxiv.org/abs/1306.5756} {arXiv:1306.5756 [hep-th]} \BibitemShut {NoStop}%
\bibitem [{\citenamefont {Altamirano}\ \emph {et~al.}(2014)\citenamefont {Altamirano}, \citenamefont {Kubiz\v{n}\'ak}, \citenamefont {Mann},\ and\ \citenamefont {Sherkatghanad}}]{Altamirano:2013uqa}%
  \BibitemOpen
  \bibfield  {author} {\bibinfo {author} {\bibfnamefont {N.}~\bibnamefont {Altamirano}}, \bibinfo {author} {\bibfnamefont {D.}~\bibnamefont {Kubiz\v{n}\'ak}}, \bibinfo {author} {\bibfnamefont {R.~B.}\ \bibnamefont {Mann}}, \ and\ \bibinfo {author} {\bibfnamefont {Z.}~\bibnamefont {Sherkatghanad}},\ }\href {\doibase 10.1088/0264-9381/31/4/042001} {\bibfield  {journal} {\bibinfo  {journal} {Class. Quant. Grav.}\ }\textbf {\bibinfo {volume} {31}},\ \bibinfo {pages} {042001} (\bibinfo {year} {2014})},\ \Eprint {http://arxiv.org/abs/1308.2672} {arXiv:1308.2672 [hep-th]} \BibitemShut {NoStop}%
\bibitem [{\citenamefont {Emparan}(2006)}]{Emparan:2006ni}%
  \BibitemOpen
  \bibfield  {author} {\bibinfo {author} {\bibfnamefont {R.}~\bibnamefont {Emparan}},\ }\href {\doibase 10.1088/1126-6708/2006/06/012} {\bibfield  {journal} {\bibinfo  {journal} {JHEP}\ }\textbf {\bibinfo {volume} {06}},\ \bibinfo {pages} {012} (\bibinfo {year} {2006})},\ \Eprint {http://arxiv.org/abs/hep-th/0603081} {arXiv:hep-th/0603081} \BibitemShut {NoStop}%
\end{thebibliography}%

\end{document}